\documentclass{iau-JDSS}
\usepackage{graphicx}

\title[short title of paper] 
{The Masses of Late-Type WN Stars}

\author[short author list]   
{G. Gr\"afener$^1$ \and W.-R. Hamann$^1$}

\affiliation{$^1$Department of Physics, University of Potsdam, 14469 Postdam, Germany}
\pubyear{2006}
\volume{Volume 14}  
\date{?? and in revised form ??}
\jname{Highlights of Astronomy, Volume 14}
\editors{K.A. van der Hucht, ed.}
\begin{document}

\maketitle

\begin{abstract}
  We present recent results for galactic WNL stars, obtained with the new
  Potsdam Wolf-Rayet (PoWR) hydrodynamic model atmospheres. Based on a
  combination of stellar wind modeling and spectral analysis we identify the
  galactic WNL subtypes as a group of extremely luminous stars close to the
  Eddington limit. Their luminosities imply progenitor masses around
  120\,$M_\odot$ or even above, making them the direct descendants of the most
  massive stars in the galaxy. Because of the proximity to the Eddington limit
  our models are very sensitive to the L/M ratio, thus allowing for a direct
  estimate of the present masses of these objects.  \keywords{stars: mass
    loss, stars: winds, outflows, stars: Wolf-Rayet.}
\end{abstract}


In the recent re-analysis of the galactic WN sample with line-blanketed
atmosphere models (\cite{ham1:06}) the WN stars turned out to form two
distinct groups in the HR diagram, which are divided by their luminosities.
Among these, the H-rich WNL\,stars, with luminosities above $10^6\,L_\odot$,
are found to the right of the ZAMS, whereas early to intermediate subtypes
show lower luminosities and hotter temperatures. The relatively large number
of extremely luminous WNL stars already implies that many of these objects
might be very massive stars in the phase of central H-burning.

Our hydrodynamic atmosphere models, on the other hand, imply that the
formation of WR-type stellar winds is caused by the proximity to the Eddington
limit (\cite{gra1:06}; \cite{gra1:05}). In fact, our models reveal a rather
strong dependence of the WR mass loss rates on the Eddington factor
$\Gamma_{\rm e}$ or, equivalently, on the $M/L$ ratio.  Weak-lined WNL stars,
with their relatively low mass loss rates, thus should have considerably
higher $M/L$ ratios than their strong-lined counterparts.

Detailed spectral modeling of weak-lined WNL stars in Carina OB\,1 indeed
indicates very high masses for these objects. Note, however, that the results
depend on the adopted distance.  For WR\,22 (WN\,7h) we find a luminosity of
$10^{6.3}\,L_\odot$ (for $m-M=12.1$) and a mass of $78\,M_\odot$, in agreement
with the mass estimate by \cite{rau1:96}. For WR\,25 (WN\,6ha) we determine
values between $110\,M_\odot$/$10^{6.4}\,L_\odot$ (for $m-M=11.8$), and
$210\,M_\odot$/$10^{6.7}\,L_\odot$ (for $m-M=12.55$). These masses are in
agreement with H-burning stars in a late pase of their main-sequence
evolution. Our models thus suggest an evolutionary sequence of the form \,O
$\rightarrow$ WNL $\rightarrow$ LBV $\rightarrow$ WN $\rightarrow$ WC\, for very
massive stars. Interestingly, WR\,25 is the only evolved object in the young
OB cluster Tr\,16, apart from the LBV prototype $\eta$\,Car. Its location at
the top of the main-sequence of this cluster, with a slightly lower luminosity
than $\eta$\,Car (see \cite{hil1:01}), strongly supports its evolutionary
stage as an LBV progenitor.

\end{document}